\documentclass[twocolumn,showpacs,preprintnumbers,amsmath,amssymb]{revtex4}

\usepackage{epsfig}
\usepackage{graphicx}
\usepackage{dcolumn}
\usepackage{bm}

\begin{document}

\preprint{Draft Revision 1.2}

\title{An efficient and accurate quantum algorithm for the Dirac equation}

\author{Jeffrey Yepez}%
\email{Jeffrey.Yepez@hanscom.af.mil}
\homepage{http://qubit.plh.af.mil}
\affiliation{
Air Force Research Laboratory \\
29 Randolph Road, Hanscom Field, Massachusetts 01731}

\date{March 25, 2002}

\begin{abstract}
An efficient quantum algorithm for the many-body three-dimensional Dirac equation is presented.  Its computational complexity is dominantly linear in the number of qubits used to spatially resolve the 4-spinor wave function.
\end{abstract}

\maketitle

The first quantum algorithm to compute a path integral was introduced by Feynman in 1965. At that time he did not refer to it as such since it was not until in 1982 that he proposed the idea of using a quantum computer to efficiently simulate quantum mechanical many-body dynamics \cite{feynman-collection}.  In the second chapter of his manuscript on path integrals published with Hibbs \cite{feynman-65-1st-qlga},  the problem is given to prove that the one-dimensional (1D) Dirac equation can be modeled by summing over all the possible zigzag paths of a single-speed particle traveling at the speed of light in a discrete two-dimensional space-time hopping from lattice site to lattice site and flipping left or right according to a deterministic local interaction rule.  The amplitude a particular path contributes to the kernel is proportional to the number of its ``collisions'' or corners.   In this way, the time evolution of the 2-spinor field of a single quantum particle is modeled by a ``gas'' of particles computing all paths simultaneously in a time-explicit fashion.  This discrete and parallelized process of local collisions and the streaming along the lattice is described by a continuous effective field theory, the 1D Dirac equation, in the limit of the infinite lattice resolution.  A solution to Feynman's ``quantum lattice gas'' problem was published in 1984 by Jacobson and Schulman \cite{jacobson-jpamg84}.    Here we give a three-dimensional (3D) solution.

In 1994, Bialynicki-Birula proposed a discrete model of the 3D Dirac equation implemented on a body-centered cubic lattice  \cite{bialynicki-birula-prd94}.  However, this model is 1st-order convergent (doubling the grid resolution merely doubles the numerical accuracy), problematic when modeling particle dynamics in an external potential.  Although the model is unitary, it is specified using non-unitary matrices and requires ad hoc lattice partitioning if implemented in parallel.  Furthermore, Bialynicki-Birula addresses only the one-body problem. Meyer published a subequent series of papers on the 1D quantum lattice gas algorithm, equivalent to Bialynicki-Birula's algorithm, and cast in the form of Feynman's original model \cite{meyer-collection}.  Meyer contributed 1D one-body numerical simulations and addressed the non-interacting lattice or checker-board problem using an additional rest particle. Yet he too did not address the many-body case nor the low-order numerical convergence issue.

Contemporaneously with Meyer, Succi published a series of papers on this subject emphasing the analogy between quantum mechanics and fluid mechanics:  the connection between the Dirac equation and the Schroedinger equation to that between the kinetic Boltzmann equation and the Navier-Stokes equation of hydrodynamics \cite{succi-collection}.   Succi's quantum lattice gas model on a cubic lattice for the 3D Dirac equation has,  at the ``kinetic'' level, the particles undergoing mixing during free propagation and is again similar to Bialynicki-Birula's model. Succi discusses the many-body case, but his algorithm runs into an ``exponential complexity wall'' \cite{succi-cpc02}.

Our quantum lattice-gas algorithm for the 3D Dirac equation is suited to direct implementation on a quantum computer using only two-qubit quantum gates and efficiently handles the many-body problem.  For pedagoical purposes, first we state the simplest quantum lattice-gas algorithm on a cubic lattice. Then, we introduce an improved version that remedies two difficiencies: the checkboard problem of non-interacting sublattices and the low-order convergence.  Finally, we recast our quantum algorithm to handle the many-body case in a second-quantized representation.

The relativistic quantum mechanical wave equation for a free particle is the linear Dirac equation
\begin{equation}
\label{dirac-equation}
\partial_t \psi = c \sum_i \alpha_i \partial_i \psi + i  \frac{m c^2}{\hbar}\beta\psi,
\end{equation}
where $\psi$ is a 4-spinor and the matrices $\alpha_i$ and $\beta$ satisfy the constraints $\alpha_i^2 = 1$, $\beta^2 = 1$, $\{\alpha_i, \alpha_j\}=0$, and $\{\beta, \alpha_i\} = 0$ so that (\ref{dirac-equation}) is equivalent to the Klein-Gordon wave equation.  Since the $2\times 2$ Pauli matrices $\sigma_i$ for $i=x,y,z$ satisfy these constraints, we can express $\alpha_i$ and $\beta$ as tensor products $\alpha_i = a\otimes \sigma_i$ and  $\beta = b\otimes {\bf 1}$,  where $a$, $b$, can be any two different Pauli matrices and where ${\bf 1}$ is the $2\times 2$ identity matrix.  We choose $a=\sigma_z$ and $b=\sigma_x$.  Then, the Dirac equation is
\begin{equation}
\label{dirac-equation-specific-form1}
\partial_t \psi = c \sum_i \sigma_z\otimes\sigma_i  \partial_i \psi + i \sigma_x\otimes 1 \frac{m c^2}{\hbar} \psi.
\end{equation}
 With the wave function defined on an infinite resolution cubical lattice at times separated by an infinitesimal duration  $\delta t$ with the grid cell size the infinitesimal length $\delta r \equiv c \delta t$, the Heisenberg evolution 
\begin{equation}
\label{heisenberg-form}
\psi' = \psi + \delta \psi= e^{\Sigma_i \sigma_z\otimes\sigma_i \delta r\partial_i -i\frac{mc^2}{\hbar}\delta t \sigma_x\otimes{\bf 1}}\psi
\end{equation}
 corresponds exactly to (\ref{dirac-equation}) in the relativistic limit where $\hbar\omega\sim mc^2$ and $\hbar k \sim mc$.  

The $4\times 4$ matrix
\begin{equation}
\sigma_z\otimes\sigma_z=
\left(
\begin{matrix}
1 & 0 & 0 & 0 \cr
0 & -1 & 0 & 0 \cr
0 & 0 & -1 & 0 \cr
0 & 0 & 0 & 1
\end{matrix}
\right), 
\end{equation}
operating with the $z$ spatial derivative in (\ref{heisenberg-form}) is diagonal whereas the matrices $\sigma_z\otimes\sigma_x$ and $\sigma_z\otimes\sigma_y$ for the $x$ and $y$ partial derivatives, respectively, are not diagonal. 

 We would like to transform (\ref{heisenberg-form}) in such a way that all the matrices operating with the spatial partial derivatives are diagonal (and hence correspond to infinitesimal shifting along the orthogonal lattice directions).  To do this, we need the two identities:
\begin{equation}
\label{similarity-transformations}
e^{-i \frac{\pi}{4}\sigma_x} e^{\varepsilon \sigma_z}e^{i \frac{\pi}{4}\sigma_x} =  e^{\varepsilon \sigma_y} 
\hspace{0.2in}
e^{i \frac{\pi}{4}\sigma_y} e^{\varepsilon \sigma_z}e^{-i \frac{\pi}{4}\sigma_x} =  e^{\varepsilon \sigma_x},
\end{equation}
that follow from $e^{i\frac{\pi}{4}\sigma_i}=\frac{1}{\sqrt{2}}\left(1+i\sigma_i\right)$ provided $\varepsilon$ is infinitesimal.  Then, using the identity ${\bf 1}\otimes e^{i\theta a} = e^{i\theta{\bf 1}\otimes a}$, the 2-spinor similarity transformations (\ref{similarity-transformations}) can be generalized to 4-spinor transformatons
\begin{eqnarray}
\label{similarity-transformations-4-spinor}
\left({\bf 1}\otimes e^{-i \frac{\pi}{4}\sigma_x} \right)e^{\varepsilon\sigma_z\otimes \sigma_z}\left({\bf 1}\otimes e^{i \frac{\pi}{4}\sigma_x} \right)& =&  e^{\varepsilon  \sigma_z\otimes\sigma_y} \\
\nonumber
\left({\bf 1}\otimes e^{i \frac{\pi}{4}\sigma_y} \right)e^{\varepsilon \sigma_z\otimes\sigma_z}\left({\bf 1}\otimes e^{-i \frac{\pi}{4}\sigma_y}\right)& =&  e^{\varepsilon \sigma_z\otimes\sigma_x},
\end{eqnarray}
which we will use to diagonalize the $x$ and $y$ spatial derivative operators in (\ref{heisenberg-form}).  Using (\ref{similarity-transformations-4-spinor}) and defining  
\begin{equation}
\label{collision-op-x1}
X^{(1)}_{\theta} \equiv e^{ i\theta\sigma_x}\otimes{\bf 1}=
\left(
\begin{matrix}
\cos\theta& 0 & i \sin\theta & 0 \cr
0 & \cos\theta & 0 &  i \sin\theta \cr
 i \sin\theta & 0 & \cos\theta & 0 \cr
0 &  i \sin\theta & 0 & \cos\theta
\end{matrix}
\right), 
\end{equation}
\begin{equation}
\label{collision-op-1x}
X^{(2)}_{\theta} \equiv{\bf 1}\otimes e^{i \theta\sigma_x}= \left(
\begin{matrix}
\cos\theta & i\sin\theta & 0 & 0 \cr
i\sin\theta & \cos\theta & 0 & 0 \cr
0 & 0 & \cos\theta & i\sin\theta \cr
0 & 0 & i\sin\theta & \cos\theta
\end{matrix}
\right), 
\end{equation}
and
\begin{equation}
\label{collision-op-1y}
Y^{(2)}_{\theta} \equiv {\bf 1}\otimes e^{i \theta\sigma_y}=\left(
\begin{matrix}
\cos\theta & \sin\theta & 0 & 0 \cr
-\sin\theta & \cos\theta & 0 & 0 \cr
0 & 0 & \cos\theta & \sin\theta \cr
0 & 0 & -\sin\theta & \cos\theta
\end{matrix}
\right), 
\end{equation}
and $S_i \equiv e^{\sigma_z\otimes\sigma_z \delta r\partial_i}$, the spatial displacement operators in the Heisenberg representation of the evolution equation (\ref{heisenberg-form}) can be written
\begin{eqnarray}
\nonumber
e^{\sigma_z\otimes\sigma_x \delta r\partial_x} &= &Y^{(2)}_{\frac{\pi}{4}}  S_x Y^{(2)\dagger}_{\frac{\pi}{4}} 
\hspace{0.25in}
e^{\sigma_z\otimes\sigma_y \delta r\partial_y} = X^{(2)\dagger}_{\frac{\pi}{4}}  S_y X^{(2)}_{\frac{\pi}{4}}\\
\label{streaming-operators}
&&e^{\sigma_z\otimes\sigma_z \delta r\partial_x} = S_z ,
\end{eqnarray}
so the evolution equation itself can be rewritten as
\begin{equation}
\label{yepez-form}
\psi' = Y^{(2)}_{\frac{\pi}{4}}  S_x Y^{(2)\dagger}_{\frac{\pi}{4}}  X^{(2)\dagger}_{\frac{\pi}{4}} S_y X^{(2)}_{\frac{\pi}{4}} S_z X^{(1)\dagger}_{\frac{mc^2\delta t}{\hbar}}\psi.
\end{equation}
This has the form of a quantum lattice-gas algorithm with local interaction (``collision'') operators $X^{(2)}_{\frac{\pi}{4}}$, $Y^{(2)}_{\frac{\pi}{4}}$ and $X^{(1)\dagger}_{\frac{mc^2\delta t}{\hbar}}$, as well as lattice-directed displacement (``streaming'') operators $S_x$, $S_y$, and $S_z$.

For numerical purposes, we would like to represent the wave function on a finite resolution grid with cell size $\delta r\rightarrow\Delta r$ and update time $\delta t\rightarrow\Delta t$. In this approximation, the wave function becomes a discrete field that exists only at the spacetime grid points $\vec x_\ell$ for $\ell=1\dots L^3$ and $t_n$ for $n=0,1,2,\dots$
\begin{equation}
\psi(\vec x_\ell, t_n)=
\left(
\begin{matrix}
\alpha (\vec x_\ell, t_n)\cr
\beta (\vec x_\ell, t_n) \cr
\mu (\vec x_\ell, t_n) \cr
\nu (\vec x_\ell, t_n)
\end{matrix}
\right)
\end{equation}
and the operators $S_i$ for $i=x,y,$ or $z$ induce a finite displacement $S_i\psi(\vec x) \rightarrow \psi(\vec x_\ell + \sigma_z\otimes\sigma_z \Delta r \hat x_i)$ of the components of the 4-spinor only along lattice directions:
\begin{equation}
\label{streaming-operator-x}
S_x\psi(x_\ell,y_\ell,z_\ell) =
\left(
\begin{matrix}
\alpha (x_\ell+\Delta r, y_\ell, z_\ell)\cr
\beta (x_\ell-\Delta r, y_\ell, z_\ell) \cr
\mu (x_\ell-\Delta r, y_\ell,z_\ell) \cr
\nu (x_\ell+\Delta r,y_\ell,z_\ell)
\end{matrix}
\right),
\end{equation}
\begin{equation}
\label{streaming-operator-y}
S_y\psi(x_\ell,y_\ell,z_\ell) =
\left(
\begin{matrix}
\alpha (x_\ell,y_\ell+\Delta r,  z_\ell)\cr
\beta ( x_\ell,y_\ell-\Delta r,   z_\ell) \cr
\mu (   x_\ell,y_\ell-\Delta r,    z_\ell) \cr
\nu (   x_\ell,y_\ell+\Delta r,    z_\ell)
\end{matrix}
\right),
\end{equation}
and
\begin{equation}
\label{streaming-operator-z}
S_z\psi(x_\ell,y_\ell,z_\ell) =
\left(
\begin{matrix}
\alpha (x_\ell, y_\ell,z_\ell+\Delta r )\cr
\beta ( x_\ell, y_\ell,z_\ell-\Delta r ) \cr
\mu (   x_\ell, y_\ell,z_\ell-\Delta r ) \cr
\nu (   x_\ell, y_\ell,z_\ell+\Delta r )
\end{matrix}
\right).
\end{equation}
These streaming operators are classical data shifting operators causing global permutations of the components of the 4-spinor wave function across the lattice and on a quantum computer can be implemented by 2-qubit local swap operators \cite{yepez-pre99}.
The collision operators
act independently on each node of the lattice and cause local quantum entanglement between component pairs of the 4-spinor.  The streaming operators in turn propagate this local on-site entanglement to next nearest neighbors so that eventually quantum entanglement covers the entire lattice.

It is possible to rewrite (\ref{yepez-form}) as a finite difference equation on a body-centered cubical lattice.  The resulting set of coupled finite difference equations are similar to the finite difference representation of the 3D Dirac equation given by Bialynicki-Birula in 1994 \cite{bialynicki-birula-prd94}.  A drawback of expressing the algorithm as a finite-difference equation is its unsuitability for a quantum computer implementation using two-qubit quantum gates whereas our manifestly unitary expression (\ref{yepez-form}) is suitable.

A continuous effective field theory for $\psi= (\alpha, \beta, \mu, \nu)$ follows in the continuum limit of the emergent finite-difference equations by Taylor expanding in $\Delta r \equiv x_{\ell+1} - x_\ell$ and in $\Delta t \equiv t_{n+1}-t_{n}$.  We obtain 
\begin{eqnarray}
\label{effective-field-theory}
\partial_t\left(\begin{matrix}
\alpha \cr
\beta \cr
\mu \cr
\nu
\end{matrix}\right)
& =&
\frac{\Delta r}{\Delta t}\partial_x\left(\begin{matrix}
-\beta \cr
-\alpha \cr
\nu \cr
\mu
\end{matrix}\right)
+
i \frac{\Delta r}{\Delta t}\partial_y\left(\begin{matrix}
\beta \cr
-\alpha \cr
-\nu \cr
\mu
\end{matrix}\right)
\\
\nonumber
&&+ \frac{\Delta r}{\Delta t}\partial_z\left(\begin{matrix}
\alpha \cr
-\beta \cr
-\mu \cr
\nu
\end{matrix}\right)
+ i\frac{m c^2}{\hbar}\left(\begin{matrix}
\mu \cr
\nu \cr
\alpha \cr
\beta
\end{matrix}\right)
+{\cal O}(c\Delta r, \Delta t),
\end{eqnarray}
which is exactly the Dirac equation (\ref{dirac-equation}) when $\Delta t  \sim \Delta r \sim \varepsilon$ are infinitesimal  
 and when the partial derivative with respect to time is defined as $\partial_t\psi \equiv \frac{\psi'-\psi}{\Delta t}$.
(\ref{yepez-form}) gives rise to perfectly unitary evolution of the discretized wave function and, therefore,  is an unconditionally stable numerical algorithm.  
The effective field theory (\ref{effective-field-theory}) is 1st-order convergent in space.

Our basic approach to improve the accuracy of the quantum algorithm is to set the grid size $\Delta r$ to be smaller than the Compton wavelength $\lambda = \frac{h}{m c}$ of the modeled particle
\begin{equation}
\label{lattice-cell-size}
\Delta r \sim  \varepsilon \frac{h}{m c} ,
\end{equation}
and to introduce a small temporal scale that is much smaller than $\frac{\lambda}{c}$
\begin{equation}
\label{lattice-time-unit}
\Delta t \sim  \varepsilon^2 \frac{h}{m c^2} .
\end{equation}
The diffusive ordering condition of spatial and temporal fluctuations typical of random walk processes, $\Delta r^2=\nu\Delta t$, provides a context to understand the scaling behavior of the small parameter $\varepsilon$. 
According to (\ref{lattice-cell-size}) and (\ref{lattice-time-unit}),  the diffusive transport coefficient is $\nu = \frac{h}{\varepsilon m}$ and the particle velocity is $\frac{\Delta r}{\Delta t} = \frac{c}{\varepsilon}$, which approaches infinity as $\varepsilon\rightarrow 0$.  In this limit, the velocity of the modeled quantum particle is relatively small, hence the resulting effective field theory should correspond to the non-relativistic limit of the Dirac equation as $\varepsilon\rightarrow 0$.

To diagonalize the streaming operators in (\ref{streaming-operators}), we used a fixed and finite rotation angle  $\frac{\pi}{4}$ independent of the grid resolution.  We will now diagonalize the streaming operators using a small rotation angle proportional to $\Delta t$.  By (\ref{lattice-time-unit}), the rotation angle is $\theta = \frac{m c^2\Delta t }{h}=\varepsilon^2$, which is dependent on the grid resolution.  The displacement operators in the Dirac equation (\ref{dirac-equation}) can be represented by interleaving streaming and collision operators on a cubical lattice as follows:
\begin{equation}
\label{displacement-operator-x}
e^{\sigma_z\otimes\sigma_x \delta r\partial_x} \rightarrow  E_x \equiv 
S_{-x}^{2,4} 
Y_{\frac{\varepsilon}{2}}^{(2)} 
S_{x}^{2,4} 
Y_{\frac{\varepsilon}{2}}^{(2)\dagger} 
S_{x}^{1,3}
Y_{\frac{\varepsilon}{2}}^{(2)} S_{-x}^{1,3} 
Y_{\frac{\varepsilon}{2}}^{(2)\dagger}
\end{equation}
and
\begin{equation}
\label{displacement-operator-y}
e^{\sigma_z\otimes\sigma_y \delta r\partial_y}\rightarrow  E_y\equiv S_{-y}^{2,4} X_{\frac{\varepsilon}{2}}^{(2)\dagger} S_{y}^{2,4} X_{\frac{\varepsilon}{2}}^{(2)} S_{y}^{1,3}X_{\frac{\varepsilon}{2}}^{(2)\dagger} S_{-y}^{1,3} X_{\frac{\varepsilon}{2}}^{(2)},
\end{equation}
where the superscripts on the streaming operators refer to individual components of the 4-spinor.  The streaming operators $S_i = S_{-i}^{2,3}S_i^{1,4}$ in (\ref{streaming-operator-x}) and (\ref{streaming-operator-y}) are now separated by collision operators.  This kind of interleaving of streaming and collision operators  removes the spurious check-board effect of independent sublattice dynamics that otherwise occurs \cite{yepez-ijmpc00b,yepez-cpc01}.  So far we treated the non-diagonal operators $e^{\sigma_z\otimes\sigma_x\delta r\partial_x}$ and  $e^{\sigma_z\otimes\sigma_y\delta r\partial_y}$, but not the displacement operator $e^{\sigma_z\otimes\sigma_z\delta r\partial_z}$ because no such improvement exists since it is diagonal.  However, if instead of using the Dirac matrix $\sigma_z\otimes\sigma_z$, we use an alternative non-diagonal representation for the $z$-direction partial derivative, then we can again employ interleaving.  Therefore, we consider this alternate form of the Dirac equation 
\begin{equation}
\label{dirac-equation-alternate}
\partial_t \psi = c \sigma_z\otimes\sigma_x  \partial_x \psi 
+ c \sigma_z\otimes\sigma_y  \partial_y \psi
+c  \sigma_y\otimes{\bf 1}  \partial_z \psi
+ i \sigma_x\otimes 1 \frac{m c^2}{\hbar} \psi.
\end{equation}
Now the displacement operator in (\ref{dirac-equation-alternate}) for the $z$-direction can be re-expressed in a  fashion similar to  (\ref{displacement-operator-x}) and (\ref{displacement-operator-y}) as
\begin{equation}
\label{displacement-operator-z}
e^{\sigma_y\otimes{\bf 1} \delta r\partial_z}\rightarrow  E_z \equiv S_z^{2,3} X_{\frac{\varepsilon}{2}}^{(1)} S_{-z}^{2,3} X_{\frac{\varepsilon}{2}}^{(1)\dagger} S_{z}^{1,4}X_{\frac{\varepsilon}{2}}^{(1)} S_{-z}^{1,4} X_{\frac{\varepsilon}{2}}^{(1)\dagger}.
\end{equation}
Then  instead of (\ref{streaming-operators}), we use (\ref{displacement-operator-x}), (\ref{displacement-operator-y}), and (\ref{displacement-operator-z}) for an improved quantum algorithm 
\begin{equation}
\label{yepez-form-interleaved}
\psi(t+\Delta t)=E_x E_y E_z X_\varepsilon^{(1)\dagger}\psi(t).
\end{equation}
In (\ref{yepez-form-interleaved}) we have appended a collision operator $X_\varepsilon^{(1)\dagger}$ to produce the mass term in the Dirac equation.

\begin{figure}[htbp]
\begin{center}
\epsfxsize=3.0in
\epsffile{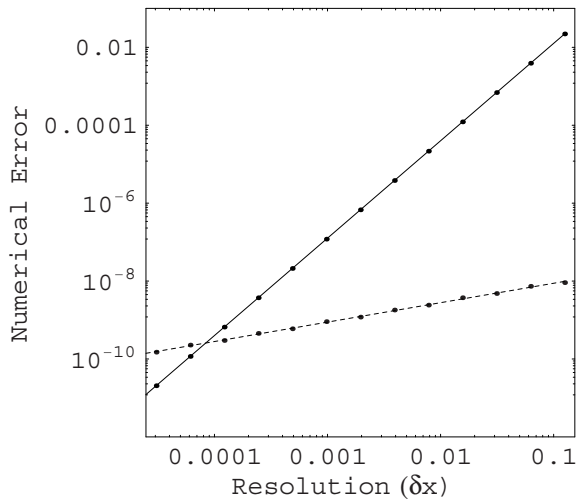}
\end{center}
\caption{\footnotesize L2 norm error $\sqrt{\frac{1}{L}\sum_{i=1}^{L}[|\psi(x_i)|^2 -|\psi_{\hbox{ex}}(x_i)|^2]}$ plotted versus grid resolution $\delta x = \frac{1}{L}$ for numerical simulations with lattice sizes from $L=8$ to $L=32768$.  The error curve's slope of the original and improved algorithm is $0.5$ (dashed line) and $2.5$ (solid line), respectively.  This demonstrates the high numerical accuracy of the improved quantum algorithm.}
\label{L2Norm}
\end{figure}

It is possible to derive a finite-difference equation representation of the quantum lattice-gas algorithm (\ref{yepez-form-interleaved}) by carrying out all the collision and streaming operations symbolically.  The result is no longer expressible on the body-centered cubic lattice.  Nevertheless, once again, a continuous effective field theory for $\psi= (\alpha, \beta, \mu, \nu)$ follows in the continuum limit and Taylor expanding in $\Delta r$ and in $\Delta t$: 
\begin{eqnarray}
\label{effective-field-theory-improved}
\partial_t\left(\begin{matrix}
\alpha \cr
\beta \cr
\mu \cr
\nu
\end{matrix}\right)
& =&
\frac{m c^2}{\hbar}\left[\Delta r\partial_x\left(\begin{matrix}
-\beta \cr
-\alpha \cr
\nu \cr
\mu
\end{matrix}\right)
+
i \Delta r\partial_y\left(\begin{matrix}
\beta \cr
-\alpha \cr
-\nu \cr
\mu
\end{matrix}\right)\right.
\\
\nonumber
&+& \left. i \Delta r\partial_z\left(\begin{matrix}
-\mu \cr
-\nu \cr
\alpha \cr
\beta
\end{matrix}\right)
+ i\left(\begin{matrix}
\mu \cr
\nu \cr
\alpha \cr
\beta
\end{matrix}\right)\right]
+{\cal O}( \Delta r^2,\Delta t),
\end{eqnarray}
which approximates the Dirac equation (\ref{dirac-equation-alternate}) when $\Delta t$ is small.  The effective field theory (\ref{effective-field-theory-improved}) is 1st-order convergent because of the error term ${\cal O}(\Delta t)$.  With the evolution operator $E=E_x E_y E_z$, we define the dual operator $\tilde E \equiv E_{-x}^\dagger E_{-y}^\dagger E_{-z}^\dagger$, by taking the adjoint of the collision operators and reversing the streaming directions.  Then, it is possible use a symmetrized evolution operator \cite{yepez-cpc01}
\begin{equation}
\label{yepez-form-interleaved-symmetrized}
\psi(t+\Delta t)=\tilde E E  e^{-\Delta t^2}\psi(t),
\end{equation}
which is better than 2nd-order accurate in space, as demonstrated in Fig.~\ref{L2Norm}.

Our quantum algorithm for the many-body Dirac equation can be expressed in terms of 2-qubit gates that conserve particle number acting on an initial ket with 4 qubits per lattice node, $|\Psi\rangle = \bigotimes_{r=1}^{L^3} |q_1(r)\rangle|q_2(r)\rangle|q_3(r)\rangle|q_4(r)\rangle$.  With $\hat a_\alpha^\dagger$, $\hat a_\alpha$, and $\hat n=\hat a_\alpha^\dagger \hat a_\alpha$ denoting the creation, annihilation, and number operator, respectively, of the $\alpha$th qubit ($1\le\alpha\le 4L^3$), the collision operators are 
\begin{eqnarray}
\nonumber
\hat X_{\alpha\beta} &=& {\bf 1}
 - i \sin\theta(\hat a_\alpha^\dagger \hat a_\beta + \hat a_\beta^\dagger \hat a_\alpha)
+(\cos\theta -1)(\hat n_\alpha + \hat n_\beta)\\
&&- 2\cos\theta \hat n_\alpha \hat n_\beta
\\
\nonumber
\hat Y_{\alpha\beta} &=& {\bf 1}
 + \sin\theta(\hat a_\alpha^\dagger \hat a_\beta - \hat a_\beta^\dagger \hat a_\alpha)
+(\cos\theta -1)(\hat n_\alpha + \hat n_\beta)\\
&&- 2\cos\theta \hat n_\alpha \hat n_\beta ,
\end{eqnarray}
where $\alpha$ and $\beta$ index different qubits at the same site. Then  (\ref{collision-op-x1}-\ref{collision-op-1y}) are rewritten as   $X_\theta^{(1)} \rightarrow \hat X_{13} \hat X_{24}$, $X_\theta^{(2)} \rightarrow \hat X_{12} \hat X_{34}$, and $Y_\theta^{(2)}\rightarrow \hat Y_{12}\hat Y_{34}$.    Hence, $2L^3$ applications of either $\hat X_{\alpha\beta}$ or  $\hat Y_{\alpha\beta}$ are required for a single collision step.
Streaming occurs by successive application of the interchange operator $\hat S_{\mu\nu} = {\bf 1} + \hat a_\mu^\dagger \hat a_\nu + \hat a_\nu^\dagger \hat a_\mu - \hat n_\mu + \hat n_\nu$  \cite{yepez-pre99}.
$(L-1)^3$ number of applications of $\hat S_{\mu\nu}$ ($\mu$ refers to one qubit-component at some site and $\nu$ to the same component at its neighboring site) are required to stream one qubit-component along a cubic lattice direction.  The total evolution operator $\hat E$ is the product of collision operators $\hat X$ and $\hat Y$ and streaming operators $\hat S$ corresponding to algorithm (\ref{yepez-form}) or some variant of (\ref{yepez-form-interleaved}) depending on the desired degree of numerical accuracy.  With the new ket $|\Psi'(t+\Delta t)\rangle = \hat E|\Psi(t)\rangle$, the resulting probability of finding a particle at site $\vec x$ is $P(\vec x) = \sum_{i=1}^4\langle \Psi' | \hat n_{\alpha+i} | \Psi'\rangle$, where $\alpha$ the index of the 1st qubit at $\vec x$. 

The computational complexity of one time step scales as
$C = \rho_{c} 2 L^3 + \rho_{s} (L-1)^3$, where $\rho_c$ and $\rho_s$ are the number of operations per node for collisions and streaming.  For the simplest algorithm (\ref{yepez-form}), $\rho_c=5$ and $\rho_s=12$, and for the improved algorithm (\ref{yepez-form-interleaved}),  $\rho_c=13$ and $\rho_s=24$.  Both $\rho_c$ and $\rho_s$ double when we use a symmetrized rule like (\ref{yepez-form-interleaved-symmetrized}) but are a fixed-cost overhead.   With $Q=4L^3$ qubits, the size of the Hilbert space is exponential $2^Q$, whereas the complexity $C=\frac{\rho_c}{2}Q + \rho_s[Q-\frac{3}{4} (2Q)^{\frac{2}{3}}+\frac{3}{2} (2Q)^{\frac{1}{3}}-1]$ for all the versions of our quantum algorithm is dominantly linear  in $Q$.

I would like to thank the Air Force Office of Scientific Research for supporting this work.

\bibliographystyle{apsrev}

\end{document}